\documentclass{elsart}
\usepackage{epsfig}

\begin{document}
\begin{frontmatter}

\title{First Results on $^{12}_{\;\Lambda}$C production at DA$\Phi$NE}

\centering{FINUDA Collaboration}

\author[polito]{M.~Agnello},
\author[victoria]{G.~Beer},
\author[lnf]{L.~Benussi},
\author[lnf]{M.~Bertani},
\author[korea]{H.C. Bhang},
\author[lnf]{S.~Bianco},
\author[unitos]{E.~Botta},
\author[units]{M. Bregant},
\author[unitos]{T.~Bressani\thanksref{corresponding}},
\author[unitog]{L.~Busso},
\author[infnto]{D.~Calvo},
\author[units]{P.~Camerini},
\author[enea]{M. Caponero},
\author[infnto]{P.~Cerello},
\author[uniba]{B.~Dalena},
\author[unitos]{F.~De~Mori},
\author[uniba]{G.~D'Erasmo},
\author[uniba]{D.~Di~Santo},
\author[unibo]{R. Don\`a},
\author[uniba]{D. Elia},
\author[lnf]{F.~L.~Fabbri},
\author[unitog]{D.~Faso},
\author[infnto]{A.~Feliciello},
\author[infnto]{A.~Filippi},
\author[infnpv]{V.~Filippini\thanksref{deceased}},
\author[uniba]{R. Fini},
\author[uniba]{M.~E.~Fiore},
\author[tokyo]{H.~Fujioka},
\author[lnf]{P.~Gianotti},
\author[infnts]{N.~Grion},
\author[jinr]{A.~Krasnoperov},
\author[lnf]{V.~Lucherini},
\author[uniba]{V. Lenti},
\author[infnba]{V. Manzari},
\author[unitos]{S.~Marcello},
\author[tokyo]{T.~Maruta},
\author[teheran]{N.~Mirfakhrai},
\author[cnr]{O.~Morra},
\author[kek]{T.~Nagae},
\author[triumf]{A.~Olin},
\author[riken]{H.~Outa},
\author[lnf]{E.~Pace},
\author[lnf]{M. Pallotta},
\author[uniba]{M.~Palomba},
\author[infnba]{A.~Pantaleo},
\author[infnpv]{A.~Panzarasa},
\author[infnba]{V.~Paticchio},
\author[units]{S.~Piano},
\author[lnf]{F.~Pompili},
\author[units]{R.~Rui},
\author[uniba]{G.~Simonetti},
\author[korea]{H.~So},
\author[jinr]{V. Tereshchenko},
\author[lnf]{S.~Tomassini},
\author[kek]{A. Toyoda},
\author[infnto]{R.~Wheadon},
\author[unibs]{A.~Zenoni}

\thanks[corresponding]{corresponding author. e-mail: bressani@to.infn.it; fax:
+39.011.6707325.}
\thanks[deceased]{deceased}
\address[polito]{Dip. di Fisica Politecnico di Torino, via Duca degli Abruzzi
Torino, Italy, and INFN Sez. di Torino, via P. Giuria 1 Torino, Italy}
\address[victoria]{University of Victoria, Finnerty Rd.,Victoria, Canada}
\address[lnf]{Laboratori Nazionali di Frascati dell'INFN, via E. Fermi 40
Frascati, Italy}
\address[korea]{Dep. of Physics,
Seoul National Univ., 151-742 Seoul, South Korea}
\address[unitos]{Dipartimento di Fisica Sperimentale, Universit\`a di
Torino, via P. Giuria 1 Torino, Italy, and INFN Sez. di Torino,
via P. Giuria 1 Torino, Italy}
\address[units]{Dip. di Fisica Univ. di Trieste, via Valerio 2 Trieste, Italy and INFN, Sez. di Trieste, via Valerio 2 Trieste, Italy}
\address[unitog]{Dipartimento di Fisica Generale, Universit\`a di
Torino, via P. Giuria 1 Torino, Italy, and INFN Sez. di Torino,
via P. Giuria 1 Torino, Italy}
\address[infnto]{INFN Sez. di Torino, via P. Giuria 1 Torino, Italy}
\address[enea]{ENEA, Frascati, Italy}
\address[uniba]{Dip. di Fisica Univ. di Bari, via Amendola 179 Bari, Italy and INFN Sez. di Bari, via Amendola 179 Bari, Italy }
\address[unibo]{Dipartimento di Fisica,
Universit\`{a} di Bologna, via Irnerio 46, Bologna, Italy and INFN,
Sezione di Bologna, via Irnerio 46, Bologna, Italy}
\address[infnpv]{INFN Sez. di Pavia, via Bassi 6 Pavia, Italy}
\address[tokyo]{Dep. of Physics Univ. of Tokyo, Bunkyo Tokyo 113-0033, Japan}
\address[infnts]{INFN, Sez. di Trieste, via Valerio 2 Trieste, Italy}
\address[jinr]{JINR, Dubna, Moscow region, Russia}
\address[teheran]{Dep of Physics Shahid Behesty Univ., 19834 Teheran, Iran}
\address[cnr]{INAF-IFSI Sez. di Torino, C.so Fiume, Torino, Italy
and INFN Sez. di Torino,
via P. Giuria 1 Torino, Italy}
\address[kek]{
High Energy Accelerator Research Organization (KEK), Tsukuba, Ibaraki
305-0801, Japan}
\address[triumf]{TRIUMF, 4004 Wesbrook Mall, Vancouver BC V6T 2A3, Canada}
\address[riken]{RIKEN, Wako, Saitama 351-0198, Japan}
\address[infnba]{INFN Sez. di Bari, via Amendola 179 Bari, Italy }
\address[unibs]{Dip. di Meccanica, Universit\`a di Brescia, via Valotti 9 Brescia, Italy and INFN Sez. di Pavia, via Bassi 6 Pavia, Italy}

\begin{abstract}
$\Lambda$-hypernuclei are produced and studied, with the FINUDA
spectrometer, for the first time at an $e^+e^-$ collider: DA$\Phi$NE, the
Frascati $\phi$-factory.
The slow negative kaons from $\phi(1020)$ decay are stopped
in thin (0.2 g/cm$^2$) nuclear targets, and $\Lambda$-hypernuclei formation
is detected by measuring the momentum of the outgoing $\pi^-$. A preliminary
analysis on $^{12}_{\;\Lambda}$C
shows an
energy resolution of 1.29 MeV FWHM on the hypernuclear levels, the best
obtained so far with magnetic spectrometers at hadron facilities.
Capture rates for the ground state and the
excited ones are reported, and compared with
previous experiments.

\bigskip
\noindent
{\it PACS}: 21.80.+a

\end{abstract}

\begin{keyword}
Hypernuclear spectroscopy, $\phi$-factory
\end{keyword}

\end{frontmatter}

\section{Introduction} \label{par1.0}
Even though the first hypernucleus was identified more than fifty years ago
\cite{Danysz53},
Hypernuclear Physics was systematically
studied only in the
last decade, in spite of  its great interest and discovery potential for
nuclear structure, strong and weak interactions and
possible quark effects in
nuclei.
The most recent experiments were performed at AGS (Brookhaven) \cite{chrien}
and at the 12 GeV PS (KEK) \cite{nagaeVarenna}, and
hypernuclei production was based on the
strangeness exchange $(K^-,\; \pi^-)$ reaction
 on nuclear targets with $K^-$ in flight and at rest, or
on the associated production $(\pi^+,\;K^+)$ one.

This experimental scenario led to the idea \cite{Bressani91}
of performing hypernuclear physics
experiments with a dedicated detector (FINUDA) using a source of $K^-$
different
from  traditional hadron
facilities; that is,  the $\phi$-factory  DA$\Phi$NE at the Frascati
National Laboratories of I.N.F.N., Italy \cite{Preger01}.

FINUDA (acronym for ``FIsica NUcleare a DA$\Phi$NE'') can be
considered an experiment of third generation in hypernuclear physics.
The original design of the FINUDA apparatus and, in
particular, the large angle covered for the
detection of charged and neutral
particles emitted after the formation and decay of hypernuclei,
allows for the simultaneous measurement of
observables like excitation energy spectra, lifetimes and
partial decay widths for mesonic and non-mesonic decay,
with high statistics and good energy resolution (better than 1~MeV).
Furthermore, these observables can be measured for different
targets at the same time, thus
reducing  systematic errors in comparing properties
of different hypernuclei.

The first FINUDA data taking at DA$\Phi$NE started in December 2003 and was
successfully concluded in March 2004.
In the following the first results from the experiment will be reported,
which fully
confirm the expected capability of FINUDA to perform high quality hypernuclear
physics at the DA$\Phi$NE collider.

\section{The FINUDA Experiment at DA$\Phi$NE}

DA$\Phi$NE (Double Annular $\Phi$-factory
for Nice Experiments)  consists of two
rings, one for electrons and the other for positrons,
that overlap in two straight sections where the beams collide.
The energy of each beam is  510~MeV in order to produce the $\phi$(1020)
meson in the collisions.

At the luminosity  $\mathcal{L}$=10$^{32}$cm$^{-2}$s$^{-1}$,
the $\phi$ meson is produced at a rate
$\sim 4.4\times 10^2$~s$^{-1}$.
The $\phi\rightarrow K^+K^-$ branching ratio is $\sim49\%$ and
therefore,
since the $\phi$  is produced almost at rest,
DA$\Phi$NE is a source of
$\sim 2.2\times 10^2$ ($K^+K^-$) pairs/s, collinear, background
free and of very low energy ($\sim 16$~MeV).
The low energy of the kaons
is the key-feature for performing
hypernuclear physics experiments at the
DA$\Phi$NE  $\phi$-factory.

The main idea of FINUDA~\cite{Bressani91,FINUDA93,FINUDA95} is to slow down
to rest the negative kaons from the $\phi\rightarrow K^+K^-$ decay
in thin solid targets, so as
to study the following formation and decay of
hypernuclei produced by the strangeness exchange reaction:
\begin{equation}
K^-_{stop}+~^AZ\rightarrow~^A_\Lambda Z+\pi^-
\label{eq1}
\end{equation}
where $^AZ$ indicates a target nucleus and $^A_\Lambda Z$ the produced
hypernucleus.
The method of producing hypernuclei via reaction (\ref{eq1})  was the
standard one with emulsions or bubble chambers in the sixties.
A first attempt
to use reaction (\ref{eq1}) even in a counter experiment was done in
1973; a $^{12}$C target was employed
and the overall energy resolution was $\sim$ 6 MeV FWHM
\cite{faessler73}. A
substantial experimental effort with a dedicated apparatus and on several
targets was then performed at KEK in the late eighties
\cite{TamuraTh,Tamura94}; however, the instrumental resolution
did not exceed 2.4 MeV FWHM.

The use of $K^-$'s from a  $\phi$-factory to produce hypernuclei
has several advantages when compared to the
extracted $K^-$ beams or intense $\pi^+$
beams~\cite{Hungerford04}.
First of all, the low-energy and almost monochromatic $K^-$ emitted from
$\phi$ decay can be efficiently stopped in thin targets
(0.2~g/cm$^{2}$).
At hadron machines, extracted $K^-$ beams require thick
targets (some~g/cm$^{2}$), in order to obtain
sufficient event rates. In addition, the uncertainty
on the interaction point and the energy straggling of the emitted
particles impair the achievable resolution.
The same problem occurs in hypernuclear spectroscopy performed via the more
efficient ($\pi^+,K^+$) reaction~\cite{Nagae01}.
In FINUDA, furthermore, the use of thin
targets along with the low-mass of the spectrometer tracking system permits
the detection of charged particles other than pions (mainly $p$'s and $d$'s)
with a solid angle similar to that of pions, and a threshold as low as
$\sim100$ MeV/$c$ for protons and $\sim 200$ MeV/$c$ for deuterons.
Finally, the cylindrical symmetry of the interaction region
allowed for the construction of a spectrometer of cylindrical shape with
a large solid angle which, for the detection of the $\pi^-$'s
coming from reaction~(\ref{eq1}),
is larger than $\pi$ sr, therefore
much bigger than those available at fixed target machines, typically
$\sim$~100~msr.
Such an acceptance, along with the excellent performances of DA$\Phi$NE,
enables the detection of  hypernuclei with a rate of about
80~hypernuclei/hour at $\mathcal{L}$=10$^{32}$ cm$^{-2}$s$^{-1}$ (with a
10$^{-3}$ capture rate).

\begin{figure}
\begin{center}
\includegraphics[height=17pc]{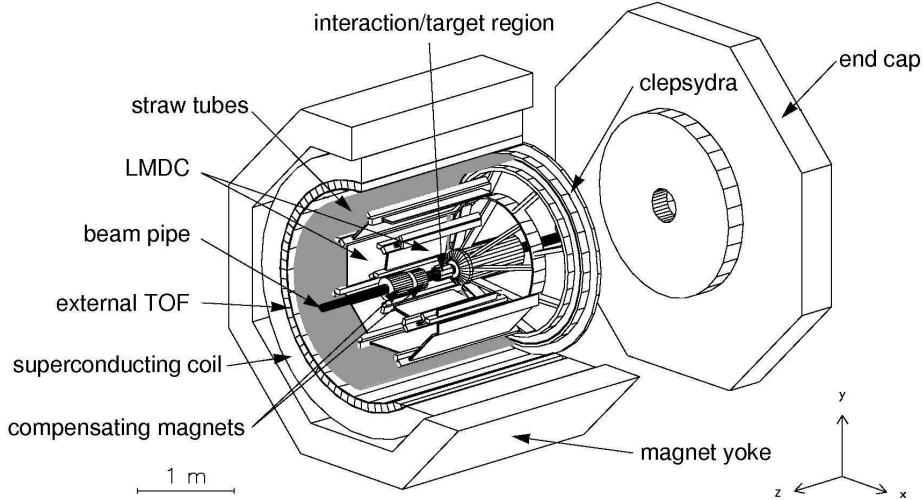}
\caption{Global view of the FINUDA detector.}
\label{fig1}
\end{center}
\end{figure}

Fig.~\ref{fig1} shows a global view of the apparatus. The layers of the
tracker are  contained
inside a superconducting solenoid, which provides a highly homogeneous
(within 2\% inside the tracking volume) magnetic
field of 1.0~T over a cylindrical volume of 146~cm radius and 211~cm length.

Three main regions can be distinguished inside the FINUDA apparatus.

\begin{figure}
\centering
\begin{tabular}{lr}
\mbox{\includegraphics[width=15pc]{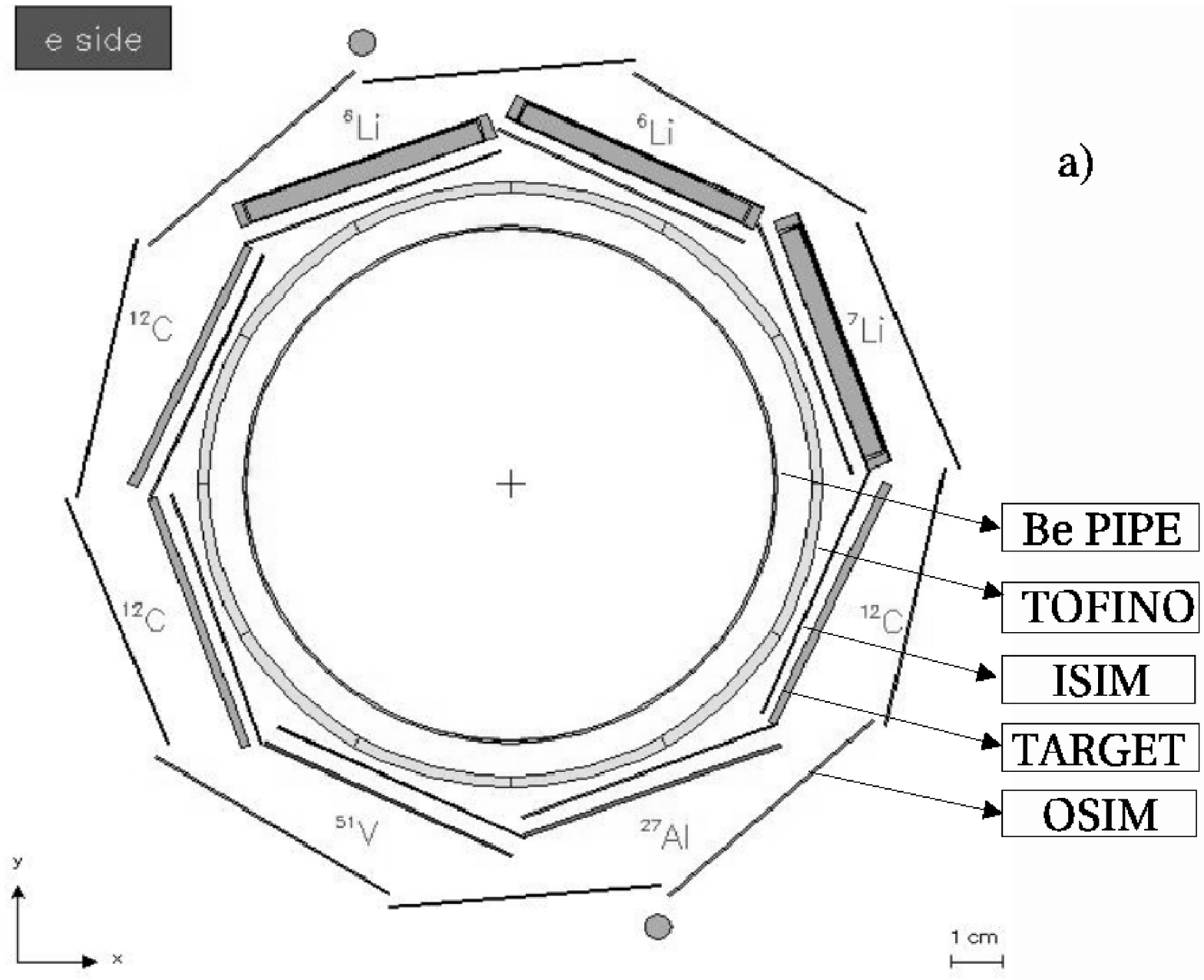}} &
\mbox{\includegraphics[width=16pc]{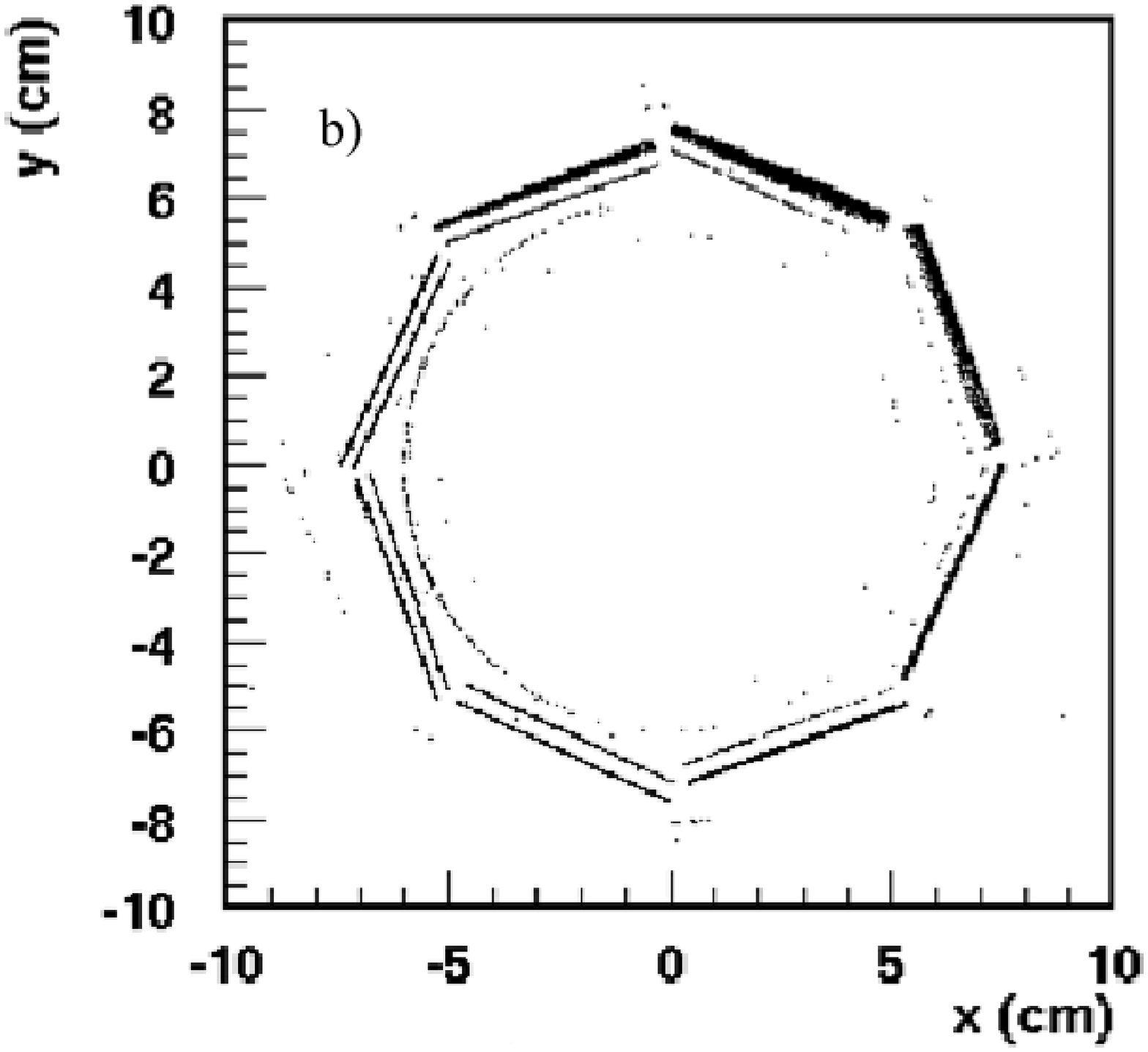}}
\end{tabular}
\caption{a) Schematic view of the interaction/target region. b)
Scatter-plot of the reconstructed $y$ vs $x$ coordinates of the $K^-$
         stopping points. See text for details.}
\label{fig2}
\end{figure}

\begin{itemize}
\item {\it The interaction/target region} is shown schematically in
      fig.~\ref{fig2}a).
      Here, the highly ionizing $(K^+,K^-)$ pairs are detected by a barrel of
      12 thin scintillator slabs (dubbed TOFINO for short), surrounding
    the beam pipe, with
      a time resolution of $\sigma \sim 250$~ps.
      The TOFINO barrel is surrounded by an octagonal array of silicon
      microstrips (ISIM) featuring a spatial resolution $\sigma \sim 30$~$\mu$m
      and an energy resolution on $\Delta E/\Delta x$ for the kaons from
      $\phi$ decay of 20\% \cite{silici}.
      Thin solid target modules are positioned at a distance of a
few millimeters
      on the external side of each element of the octagon.
      The task of the ISIM detector is the determination
      of the interaction points of the $(K^+,K^-)$ pairs in the thin
      targets.

\item {\it The external tracking device} consists
of four different layers of
       position sensitive detectors. It is arranged in
       cylindrical symmetry and is
       immersed in a He atmosphere to reduce the
       effects of the multiple Coulomb scattering.
       The trajectories of charged particles coming from the targets and
       crossing the tracking system are measured by:
      (i)~a first array of ten double-sided silicon microstrip modules (OSIM)
       placed close to
       the target elements (see fig.~\ref{fig2}a);
      (ii)~two arrays of eight planar low-mass drift chambers
      (LMDC) filled with a (70\%He-30\%C$_4$H$_{10}$) mixture, featuring a
      spatial resolution $\sigma_{\rho\phi}\sim 150$~$\mu$m and
      $\sigma_{z}\sim 1.0$~cm \cite{nimCamere};
      (iii)~a straw tube detector, composed by six layers of longitudinal
      and stereo tubes, which provide a spatial resolution
      $\sigma_{\rho\phi}\sim 150$~$\mu$m and $\sigma_{z}\sim 500$~$\mu$m
      \cite{nimStraw}.
      The straw tubes are
      positioned at 1.1 m from the beams interaction point.
      With the magnetic field set at 1.0~T,
      the design momentum resolution of the spectrometer, for
      270~MeV/c $\pi^-$'s, is
      ${\Delta p}/{p}$=0.4\%~FWHM. It corresponds to an energy resolution on
      hypernuclear spectra better than 1.0~MeV.
      On the other hand, the energy resolution, for the 80 MeV protons
      emitted in the hypernuclear non-mesonic decay, is 1.6~MeV~FWHM.

\item {\it The external time of flight barrel} (TOFONE) is composed by 72
       scintillator slabs, 10~cm thick and 255 cm long, and provides
       signals for the first level
       trigger and for the measurement of the time-of-flight of
       the charged
       particles, with a time resolution  $\sigma\sim$ 350 ps.
       Moreover, it allows for the detection of neutrons following
hypernucleus decays with an efficiency of $\sim$10\%, an angular acceptance of
       70\% and an energy resolution of 8~MeV~FWHM for neutrons of 80~MeV
        \cite{tofone}.
\end{itemize}

       Further details concerning the design and performances of the FINUDA
       apparatus can be found
    in Refs. \cite{Zenoni99,Gianotti01,Zenoni02}.

An important feature of the FINUDA apparatus (see
fig.~\ref{fig2}a) is the possibility to host eight different targets
close to  the
interaction region; therefore, the possibility of
obtaining data on different hypernuclei at the same
time.
For the starting run, the following targets were selected~\cite{Bressani03b}:
two $^6$Li (isotopically enriched to 90\%), one $^7$Li
(natural isotopic abundance), three $^{12}$C, one $^{27}$Al
and one $^{51}$V.
Physical motivations for the performed choice are described in Refs.
\cite{Bressani03b,ZenoniVarenna}.

\section{Data taking and apparatus performances}

Many experimental tests were performed during the data taking
in order to monitor
the machine performance  as well as the calibrations of the
spectrometer.
Hereafter the most relevant ones are listed:
\begin{enumerate}
\item the luminosity of the DA$\Phi$NE collider was continuously evaluated by
means of the Bhabha scattering events, and was
in agreement (within $\sim 10\%$) with the values provided by the machine.
The top luminosity reached during the run was
0.7$\times 10^{32}$ cm$^{-2}$s$^{-1}$, with a daily integrated luminosity of
about 4 pb$^{-1}$;
\item the profile of the interaction region was
also continuously monitored by FINUDA,
and used to control the collider;
\item the energy of the colliding beams
was measured on-line via Bhabha scattering and through
the reconstruction of the $K_S\rightarrow \pi^+\pi^-$ invariant mass, where the
$K_S$'s are due to the $\phi\rightarrow K_SK_L$ decay.
\end{enumerate}

The trigger selecting hypernucleus formation events requires
two fired
back-to-back TOFINO slabs, with signal amplitude
above
an energy threshold accounting for the high ionization of slow kaons, and a
fast coincidence on the TOFONE barrel~\cite{FINUDA95}.
This allows $(K^+,K^-)$ pairs, together with a fast particle crossing the
spectrometer and hitting the external scintillator barrel, to be selected
against the physical background coming from the other $\phi$
decays or
against fake events generated by the accelerator electromagnetic background.

The reconstruction procedure of the $\phi$ formation point and of the kaon
directions and momenta at vertex uses the kaon interaction  points in the ISIM
modules, identified through their high stopping power. The procedure is
based on a two helix algorithm which accounts for the kinematics of the
$\phi$ decay, the average value of the $\phi$ mass, the crossing angle (12.5
mrad) of the $e^+e^-$ beams, measured by using Bhabha events, and the geometry
of the vertex region. The stopping points of the kaons in the targets are
computed by a tracking procedure based on the GEANE package \cite{GEANE},
which performs a numerical integration of the trajectory starting from the
$\phi$ formation point and the kaon direction and momenta and accounting for
the geometrical structure and the material composition of the FINUDA
interaction region.

The beam crossing angle determines a small total momentum of the $\phi$
(boost: 12.3 MeV/$c$) directed towards the positive $x$ side. This boost adds
to the 127 MeV/$c$ average momentum of the kaons from the $\phi$ decay
introducing a left-right asymmetry clearly visible in Fig.
\ref{fig2}b), which shows the scatter-plot of the reconstructed $y$ vs $x$
coordinates of the $K^-$ stopping points.
The distribution of points on the outer octagon represents
the positions of
the eight targets, where most of the $K^-$ stop
($\sim 75\%$ of all $K^-$ interactions in the apparatus).
A partial accumulation of points also occurs
on the left-side ISIM modules (10\%).
The events corresponding
to the $K^-$'s stopping in the ISIM modules
provide an additional sample in a supplementary silicon target.
The remaining density of points partially depicts TOFINO.

Hypernuclear events are selected by the simultaneous presence of $K^+$ and
$K^-$ particles. The $K^+$'s enable the $K^-$ tagging and moreover offer
the possibility to perform an accurate and continuous in-beam calibration
of FINUDA. The positive kaons, stopping in the target array, decay at rest
with a mean life of 12.4 ns. The two main two-body decays
$K^+\rightarrow\mu^+\nu_{\mu}$ (B.R.=63.51\%) and
$K^+\rightarrow\pi^+\pi^0$ (21.16\%) are a source of monochromatic particles
fully crossing the spectrometer, with momenta
235.5~MeV/c for the $\mu^+$ and and 205.1~MeV/c for the $\pi^+$,
respectively.
The absolute scale of the momenta was determined with a precision
better than 200 keV/$c$, even in the simplified hypothesis
(applied in the analysis presented here) of a
constant magnetic field of 1.0 T, directed along the $z$ axis, in the
whole tracking volume.
This precision can be assumed as the systematic error on
the measurement of the particles' momenta in the range between 200 and
300 MeV/$c$.

\begin{figure}
\begin{center}
\includegraphics[width=15pc,height=15pc]{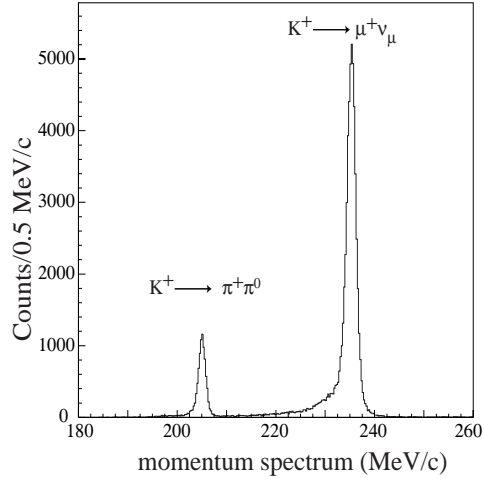}
\caption{Momentum distribution of the positive tracks coming from the
         stopped $K^+$. The peak at 236~MeV/c corresponds to the two body
     decay $K^+\rightarrow\mu^+\nu_\mu$, the peak at 205~MeV/c corresponds
     to the two body decay $K^+\rightarrow\pi^+\pi^0$.
     }
\label{fig3}
\end{center}
\end{figure}

For the present analysis
only high quality tracks were selected. Such
tracks are emitted in the forward hemisphere, with respect to the direction
of the $K^+$, and
cross a minimum amount of materials inside the spectrometer.
Fig.~\ref{fig3} shows the momentum distribution of the positive tracks
coming from the stopped $K^+$. The two peaks at 236~MeV/c and 205~MeV/c
correspond to the previously mentioned decays.
The tails on the left of the two peaks are due to different contributions, the
biggest part played by
instrumental effects due to the momentum
loss of particle crossing the edges of the chambers and their supports.
Moreover, in this region two additional $K^+$ decay channels open:
the $K^+_{e 3}$ mode
(B.R.=4.8\%), giving a continuum spectrum of positrons
(which cannot be distinguished from $\mu^+$'s) ending
at 228 MeV/$c$,  and the
$K^+_{\mu 3}$ one (3.2\%), which gives again a continuum
spectrum with end point at 215 MeV/$c$.
By analyzing these different contributions to the peaks shape
one can conclude that the asymmetry affects, overall, the gaussian line shape
at the level of about 4\%.
This peak asymmetry was however not considered in the fit of the spectra
described in Sec. \ref{analisi}, since other
error sources were overwhelming.

From the width of the $\mu^+$ peak the present momentum resolution of the
apparatus can be estimated to be
${\Delta p}/{p}$=0.6\%~FWHM, which corresponds to 1.29 MeV FWHM for the
hypernuclear levels in agreement with the results of the hypernuclear spectra
reported in the next section. We expect that the momentum resolution of the
spectrometer should improve to the design value of 0.4\% FWHM once the
final detector calibration and alignment will be performed, and the
mapped magnetic field will be inserted in the reconstruction and fitting
procedure.


\section{Results on $^{12}_{\;\Lambda}$C spectroscopy. Discussion and
conclusions.}
\label{analisi}

In order to evaluate the capabilities of FINUDA to yield relevant
spectroscopic parameters, the analysis started
from $^{12}$C targets. We recall that for
$^{12}_{\;\Lambda}$C an excitation spectrum with a
1.45 MeV FWHM resolution was recently obtained at KEK using the
$(\pi^+,\; K^+)$ reaction at 1.05 GeV/$c$ by the E369 Collaboration
\cite{Nagae01}.

The spectra out of only two of the three available
$^{12}$C targets were added since
the third one showed a slight systematic energy
displacement, of
about 0.5 MeV. The reason of this is under study, and therefore
for the current
analysis these data
are not included. The requirement of high quality tracks (long tracks crossing
the whole spectrometer, with a hit on each tracking detector, {\it i.e}
OSIM, LMDC's and straw tubes) reduced
the analysed data to about the 40\% of the whole available sample of events
with vertex coming from a $^{12}$C target.

\begin{figure}
\begin{center}
\includegraphics[width=15pc,height=15pc]{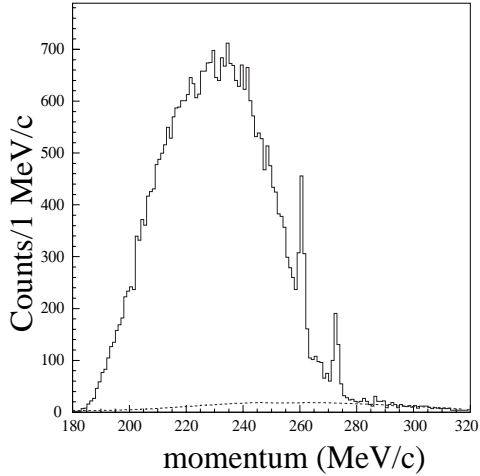}
\caption{Spectrum of the momentum of the $\pi^-$ emitted from the interaction
vertex of a $K^-$ onto a carbon target. The dashed line represents the
contribution from $K^-$ absorption by two nucleons (process d) in the text).}
\label{fig:c12spectrum}
\end{center}
\end{figure}

The raw momentum spectrum of the $\pi^-$ coming from the analysed
$^{12}$C targets
is shown in Fig. \ref{fig:c12spectrum}.
Different processes produce $\pi^-$ after $K^-$ absorption and reproduce
well the experimental spectra \cite{Tamura94}:
\begin{description}
\item{a)} quasi-free $\Sigma^+,\; \Sigma^0$ and $\Lambda$
production: $K^-p\rightarrow\Sigma^+\pi^-$,
$K^-n\rightarrow\Sigma^0\pi^-$, $K^-n\rightarrow\Lambda\pi^-$;
\item{b)} quasi-free $\Lambda$ decay: $\Lambda\rightarrow p\pi^-$;
\item{c)} quasi-free $\Sigma^-$ production: $K^-p\rightarrow\Sigma^-\pi^+$,
 followed by $\Sigma^-\rightarrow n\pi^-$;
\item{d)} two nucleon $K^-$ absorption: $K^-(NN)\rightarrow\Sigma^-N$,
followed by
$\Sigma^-\rightarrow n\pi^-$.
\end{description}

All the mentioned reactions were simulated in detail in the
FINUDA Monte Carlo program.
The simulated events were reconstructed by the same
program used for the real events, with the same selection criteria, in order
to accurately take into account the geometrical acceptance and the
reconstruction efficiency of the apparatus.
In particular, the size of the spectrometer and the value of the magnetic
field determine an acceptance momentum cut of about
180~MeV/c for four-hits tracks,
which excludes most of the reactions producing
low energy $\pi^-$'s.
However, in the momentum region where the bound states of
$^{12}$C are expected (beyond $\sim260$ MeV/$c$), only process {d)} is
contributing.
We remark that both processes {c)} and {d)} are due to $\Sigma^-$
decay in flight, but the $\pi^-$ distribution from the process {c)} is
peaked at 190 MeV/$c$, and goes to zero beyond 260 MeV/$c$.
The dashed line in Fig. \ref{fig:c12spectrum} represents the
contribution due to process {d)}, normalized to the number of entries
in the $(275\div 320)$ MeV/$c$ momentum region,
beyond the physical region for the production of $\Lambda$-hypernuclei via
reaction (\ref{eq1}).

\begin{figure}
\centering
\begin{tabular}{lr}
\mbox{\includegraphics[width=15pc,height=15pc]{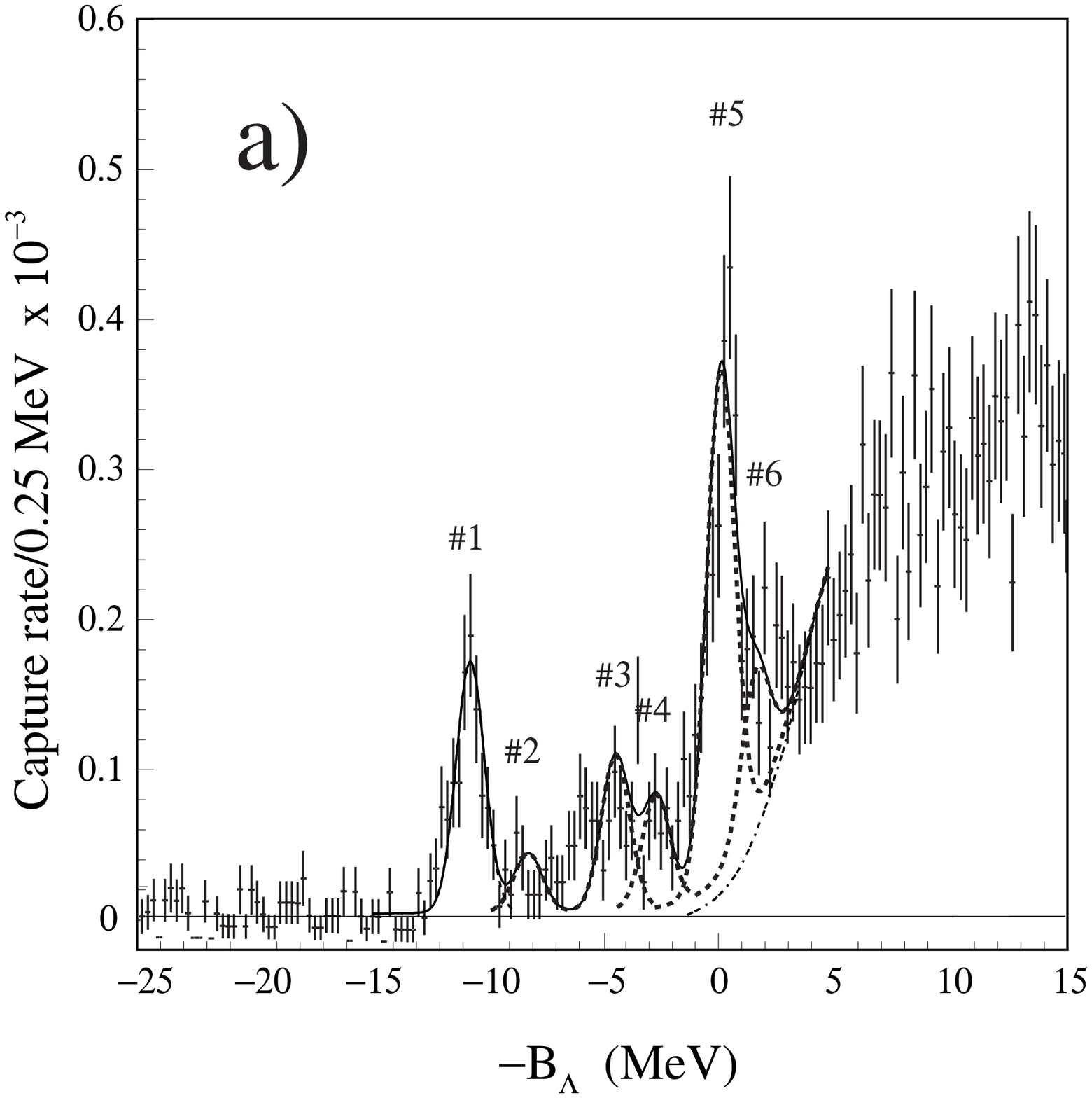}} &
\mbox{\includegraphics[width=15pc,height=15pc]{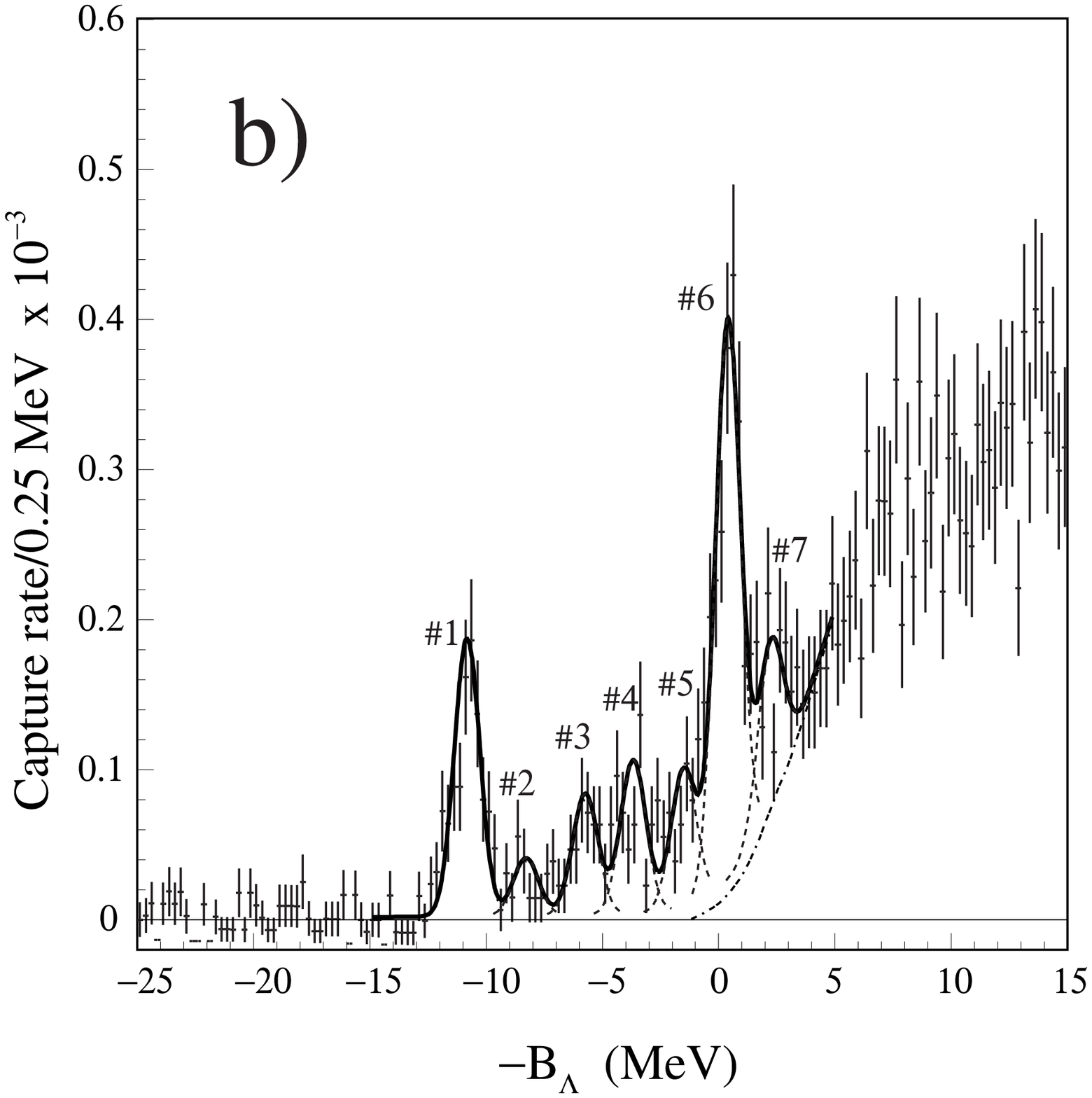}}
\end{tabular}
\caption{
$\Lambda$ binding energy spectrum of $^{12}_{\;\Lambda}$C
measured by the FINUDA experiment. a) The solid line represents the
result of a
fit with 6 gaussian functions $(\#1\div \#6)$, as explained in the text;
b) the solid line represents the result of a fit with 7 gaussian functions
as explained in the text. The dot-dashed line starting at $B_\Lambda= -1$ MeV
represents the contribution from the quasi-free $\Lambda$ production. The
dotted lines represent the result of a gaussian fit on every single peak.
     }
\label{fig:spettroC12}
\end{figure}

In order to obtain the $\Lambda$ binding energy distribution
the {d)} process is subtracted from the $\pi^-$ momentum distribution,
and the momenta are converted into binding energies ($-B_\Lambda$).
The two prominent peaks, as can be seen in Figs. \ref{fig:spettroC12}a)
and b),
at $B_\Lambda$ around 11 MeV (ground state)
and 0 MeV,
were already observed in
previous experiments \cite{faessler73,TamuraTh} and interpreted as
$(\nu p^{-1}_{\frac{1}{2}},\Lambda s)$ and
$(\nu p^{-1}_{\frac{3}{2}},\Lambda p)$ ($\nu$= nucleon).
The experimental energy
resolution was determined by fitting the $B_\Lambda\simeq$ 11 MeV
peak with a gaussian curve ($\chi^2/d.o.f. \ = \ 1.71$), 
and amounts to 1.29 MeV FWHM.
The ground state of $^{12}_{\;\Lambda}$C is assumed to be a 
single state. Indeed,
it is known that it consists of a $(1^-,\; 2^+)$ doublet, but theoretical
calculations predict splittings of 70 keV \cite{millener}, 80 keV
\cite{fetisov} and 140 keV \cite{itonaga2} between them,
one order of magnitude smaller
than the present instrumental resolution. The peak at about 0. MeV has
a more complicated structure, and we tried to disentagle different
contributions in the analysis described in the following.

The experimental spectrum closely resembles the one from E369 experiment
\cite{Nagae01}.
This is expected, as the production
of hypernuclear states is, in first approximation,
determined by the momentum transferred to $\Lambda$'s,
which is grossly comparable
for both experiments ($\sim 250$ MeV/$c$ for FINUDA, $\sim 350$ MeV/$c$
for E369). The $\sim 100$ MeV/$c$ difference may account for the different
yield of the two main peaks.

The absolute values of the capture rates for the different peaks could be
obtained in a simple way by the method of the $K^-$ tagging. Indeed, in the
events where the $K^+$ is seen to decay in the $K_{\mu 2}$ and $K_{\pi 2}$
decay modes with the produced $\mu^+$ or $\pi^+$ crossing the spectrometer
and hitting the TOFONE barrel, we are sure the trigger condition on the
 prompt TOFONE coincidence has been satisfied by the charged products of
the $K^+$ decay.

Hence, in these events triggered by the decay products of the $K^+$, the
interactions of the corresponding $K^-$ in the targets are observed without
any trigger bias. Using this subsample of events, the number of $K^-$
stopping in the targets can be counted directly and the number of $\pi^-$
produced by the $K^-$ interactions can be accurately determined by only
correcting for the apparatus acceptance for $\pi^-$ of selected momentum and
for detector efficiency. The acceptance is calculated using the FINUDA
Monte Carlo and the detector efficiency is determined by calibration data.

The value obtained for the $^{12}_{\;\Lambda}$C ground state
formation is
$(1.01\pm 0.11_{stat}\pm 0.10_{sys})\times10^{-3}$/(stopped $K^-$).
It agrees
very well with the value $(0.98\pm 0.12)\times 10^{-3}$/(stopped $K^-)$ 
measured at KEK
\cite{Tamura94}; we recall that the first generation CERN experiment reported
the value $(2\pm 1)\times 10^{-4}$/(stopped $K^-$) \cite{faessler73}.

In between the two main peaks, there are also
indications of other states produced with weaker strength.
In order to reproduce, at least qualitatively, this spectrum
six gaussian functions were used,
centered at the $B_\Lambda$ values reported in
Ref. \cite{Nagae01}; the widths were fixed, for all of them, to
$\sigma=0.55$ MeV, corresponding to the experimental resolution.
The abscissa scale is affected only by a scale error of $\pm 80$ keV.
The result of this fit
is shown in Fig. \ref{fig:spettroC12}a).

The spectrum is not well reproduced,
the resulting reduced $\chi^2/d.o.f.$ is 3.8 (for 64 {\it d.o.f.}), and
in particular the region  $-10\ {\mathrm{MeV}} <-B_\Lambda<-5\ \mathrm{MeV}$
is poorly
fitted. The
capture rates for these different contributions normalized to the ground state
capture rate are reported in the second
column of the upper part of Table
\ref{tab:tab}.
A better $\chi^2/d.o.f.$= 2.3 is obtained by adding a further
contribution, and leaving the positions of the seven levels free
(57 {\it d.o.f.}). Their
values are reported in the second column of the lower part of Tab.
\ref{tab:tab}.
The capture rates for these different contributions are again normalized to the
capture rate for the $^{12}_{\;\Lambda}$C ground state formation.
The result of the fit is shown in Fig.
\ref{fig:spettroC12}b). A contribution from the quasi-free
$\Lambda$-production, starting from $B_\Lambda=0$ and properly smeared by taking
into account the instrumental resolution, was included in both fits.

    \begin{table}[h]
      \begin{tabular}{c|c|c}
    \hline
    Peak number & $-{B_{\Lambda}~(\mathrm{MeV})}$ &
        Capture rate/(stopped $K^-$)$[\times 10^{-3}]$\\
      &  {\it (Fixed~at~E369~values)} &
           \\
    \hline
    1 & $-10.76$ & $1.01\pm 0.11_{stat}\pm 0.10_{syst}$    \\
    2 & $-8.25$  & $0.23\pm 0.05$\\
    3 & $-4.46$  & $0.62\pm 0.08$\\
    4 & $-2.70$  & $0.45\pm 0.07$\\
    5 & $-0.10$  & $2.01\pm 0.14$\\
    6 & $ 1.61$  & $0.57\pm 0.11$\\
    \hline
    \hline
    Peak number & $-{B_{\Lambda}~(\mathrm{MeV})}$ &
        Capture rate/(stopped $K^-$)$[\times 10^{-3}]$\\
    \hline
    1 & $-10.94 \pm 0.06$ & $1.01\pm 0.11_{stat}\pm 0.10_{syst}$    \\
    2 & $-8.4   \pm 0.2$  & $0.21\pm 0.05$ \\
    3 & $-5.9   \pm 0.1$  & $0.44\pm 0.07$\\
    4 & $-3.8   \pm 0.1$  & $0.56\pm 0.08$\\
    5 & $-1.6   \pm 0.2$  & $0.50\pm 0.08$\\
    6 & $0.27   \pm 0.06$ & $2.01\pm 0.17$\\
    7 & $2.1    \pm 0.2$  & $0.58\pm 0.18$\\
\hline
      \end{tabular}
      \caption{Results from $-B_{\Lambda}$ spectrum fits: the upper
part of the table
    corresponds to a fit performed with the same
peaks layout of E369 experiment \protect{\cite{Nagae01}}, with 6 gaussian
functions. The
lower part corresponds to a fit with 7 hypernuclear levels. The last column
reports the capture rates corresponding to each peak. The errors reported for
peak $(\#2\div\#6)$ in the upper part and  $(\#2\div\#7)$ in the
lower part of the table do not include the error on the $^{12}_{\;\Lambda}$C
ground state capture rate. The errors on the rates of peaks \#6 and \#7 take
into account the error on the subtracted background.}
\label{tab:tab}
    \end{table}

The peaks $\# 2$ and $\# 3$ can be attributed to the $^{11}$C core
excited
 states at 2.00 and 4.80 MeV.
 The excitation of these states was expected in several theoretical
 calculations \cite{gal,Motoba98};
their energies may be sensitive to the
  $\Lambda$-$N$ interaction matrix elements.
 However, the peak $\#$4 and a newly observed peak $\#$5 are not explained
 with such a simple way.
Excluding from the fit the peak $\#$5 the value for $\chi^2/d.o.f.$ worsened to 3.3.
There exist several positive-parity excited states of the $^{11}$C core
 in this energy region which could contribute the these structures
\cite{Motoba98}.
It can be noticed that the integrated strength for the excitation  of all
these weakly excited states compared
to that of the two main peaks
is more than twice larger than the one reported by E369.

The sum of the capture rates for the
$B_\Lambda=0.27$ MeV and $B_\Lambda=2.1$ MeV states
is $(2.59\pm 0.19_{stat})\times 10^{-3}$/(stopped $K^-)$,
and agrees with the KEK result $(2.3\pm 0.3)\times 10^{-3}$/(stopped $K^-)$ 
\cite{Tamura94},
in which the contributions for the two
states were not resolved. The CERN experiment \cite{faessler73} reports 
$(3\pm 1)\times 10^{-4}$/(stopped $K^-)$.
In the present analysis these states are indeed
resolved, though, inevitably, strongly correlated in our fit.
It is however remarkable that the relative intensities
for the contributions at
$B_\Lambda = 0.27$ MeV and
$B_\Lambda = 2.1$ MeV are close to the values
found by Dalitz {\it et al.}
\cite{dalitz86} in an emulsion experiment.

Theoretical calculations for the ground state formation quote the values
$0.33\times 10^{-3}$/(stopped $K^-$) \cite{hunger44}, 
$0.23\times 10^{-3}$/(stopped $K^-$) \cite{hunger48} and 
$0.12\times 10^{-3}$/(stop\-ped $K^-$) \cite{hunger46}. 
Analogous calculations for the capture rate leading to states in which the 
$\Lambda$ is in a $p$ state  quote \cite{ahmed}, respectively, 
$0.96\times 10^{-3}$/(stopped $K^-$) according to the theoretical
prediction of Ref. 
\cite{hunger44}, and $0.59\times 10^{-3}$/(stopped $K^-$) following Ref.
\cite{hunger46}. As general remark it may be noticed that our measured 
values are larger by factors $(3\div 6)$ as compared with theoretical
predictions. Finally, the pattern of the relative strength for the 
excited-core states is also significantly larger than the theoretical
calculation reported in Ref. \cite{itonaga90}.

In conclusion, the method of producing hypernuclei
stopping in thin nuclear targets the low energy $K^-$ from $\phi$ decay at
a $\phi$-factory was proved to work,
and may be used to perform accurate
measurements on many hypernuclei observables.
A first analysis
allowed to achieve this purpose and to obtain also some
interesting new physical information.

\section{Acknowledgements}

We are greatly indebted to Dr. S. Bertolucci, former Director of LNF, for
his continuous encouragement and help. Dr. P. Raimondi and the DA$\Phi$NE
machine staff are warmly acknowledged for their very skillful handling of the
DA$\Phi$NE collider.

The excellent and qualified contribution of the whole FINUDA technical staff,
at all stages of the experiment setting-up, is deeply appreciated.


\end{document}